\documentclass[twocolumn]{article}
\usepackage[inner=1cm,outer=1cm]{geometry}
\usepackage{authblk}

\usepackage[backref=false,maxbibnames=99,doi=false,isbn=false,url=false,maxcitenames=2,bibstyle=numeric,hyperref=auto,backend=bibtex,natbib=true,style=numeric,citestyle=numeric,sortcites=true]{biblatex}
\bibliography{crowd.bib}

\usepackage[linesnumbered,ruled,vlined]{algorithm2e}

\usepackage[utf8]{inputenc}
\usepackage{amsmath}
\usepackage{amssymb}
\usepackage{graphicx}
\usepackage{verbatim}
\usepackage{xspace}

\usepackage[usenames,dvipsnames]{color}
\usepackage[colorlinks=true,linkcolor=Mahogany,citecolor=blue]{hyperref}
\usepackage[overload]{textcase}
\preto\equation{\par\nobreak\small\noindent}
\preto\align{\par\nobreak\small\noindent}
\expandafter\preto\csname align*\endcsname{\par\nobreak\small\noindent}
\usepackage{textcomp}
\usepackage{enumitem}
\usepackage{caption}
\usepackage{soul}
\usepackage{ntheorem}
\usepackage{cleveref}
\usepackage[usenames,dvipsnames]{color}
\usepackage{color, colortbl}
\usepackage{xcolor}

\newenvironment{pf}[2]{\noindent\emph{Proof of {#1} {#2}:}\hspace*{1mm}}{ \hspace*{\fill} $\blacksquare$ }

\newenvironment{pfof}{\noindent{\em Proof:} }{ \hfill $\blacksquare$\\ }

\newcommand{\CUCB}{2D-UCB\xspace}
\newcommand{\PRIN}{Well-Behaved }
\newcommand{\CHOPT}{2D-OPT\xspace}
\newcommand{\Prob}{\mathbb{P}}

\newcommand{\squishlist}{
 \begin{list}{$ullet$}
  { \setlength{\itemsep}{0pt}
     \setlength{\parsep}{3pt}
     \setlength{\topsep}{3pt}
     \setlength{\partopsep}{0pt}
     \setlength{\leftmargin}{1.5em}
     \setlength{\labelwidth}{1em}
     \setlength{\labelsep}{0.5em} } }

\newcommand{\squishlisttwo}{
 \begin{list}{$ullet$}
  { \setlength{\itemsep}{0pt}
    \setlength{\parsep}{0pt}
    \setlength{	opsep}{0pt}
    \setlength{\partopsep}{0pt}
    \setlength{\leftmargin}{2em}
    \setlength{\labelwidth}{1.5em}
    \setlength{\labelsep}{0.5em} } }

\newcommand{\squishend}{
  \end{list}  }
\newtheorem{definition}{Definition}

\newtheorem{theorem}{Theorem}
\newtheorem{lemma}[theorem]{Lemma}

\DeclareMathOperator*{\argmax}{\arg\max}

\crefname{lemma}{lemma}{lemmas}
\Crefname{lemma}{Lemma}{Lemmas}

\title{An Optimal Bidimensional Multi-Armed Bandit Auction for Multi-unit Procurement}

\author[1]{Satyanath Bhat}
\author[2]{Shweta Jain}
\author[3]{Sujit Gujar}
\author[4]{Y Narahari}

\affil[1]{Computer Science and Automation \\
Indian Institute of Science \\
\url{satya.bhat@gmail.com}}
\affil[2]{Computer Science and Automation \\
Indian Institute of Science \\
\url{jainshweta@csa.iisc.ernet.in }}
\affil[3]{Artificial Intelligence Laboratory \\
EPFL\\
\url{sujit.gujar@epfl.ch }}
\affil[4]{Computer Science and Automation \\
Indian Institute of Science \\
\url{hari@csa.iisc.ernet.in}}

\begin{document}
\maketitle

\begin{abstract}
We study the problem of a buyer (aka auctioneer) who gains stochastic rewards by procuring
multiple units of a service or item from a pool of heterogeneous strategic agents. 
The reward obtained for a single unit from an allocated agent depends on the inherent quality of 
the agent; the agent's quality is fixed but unknown.
Each agent can only supply a limited number of units (capacity of the agent). 
The costs incurred per unit and capacities are private information of the agents.
The auctioneer is required to elicit costs as well as capacities (making
the mechanism design bidimensional) and 
further, learn the qualities of the agents as well, with a view to maximize her utility. 
Motivated by this, we design a bidimensional multi-armed bandit procurement
auction that seeks to maximize the expected utility of the auctioneer
subject to incentive compatibility and individual rationality
while simultaneously learning the unknown qualities of the agents.
We first assume that the qualities are known and propose an optimal, 
truthful mechanism \CHOPT for the auctioneer to elicit costs and  capacities. 
Next, in order to learn the qualities of the agents in addition, we provide sufficient 
conditions for a learning algorithm to be Bayesian incentive compatible and  individually 
rational.  We finally design a novel learning mechanism, \CUCB  that is stochastic 
Bayesian incentive compatible and individually rational.
\end{abstract}

% Note that the category section should be completed after reference to the ACM Computing Classification Scheme available at
% http://www.acm.org/about/class/1998/.

%
%\category{I.2.11}{Distributed Artificial Intelligence}{Intelligent agents} \category{I.2.6} {Learning}{Miscellaneous}

%A category including the fourth, optional field follows...
%\category{D.2.8}{Software Engineering}{Metrics}[complexity measures, performance measures]

%General terms should be selected from the following 16 terms: Algorithms, Management, Measurement, Documentation, Performance, Design, Economics, Reliability, Experimentation, Security, Human Factors, Standardization, Languages, Theory, Legal Aspects, Verification.
%
%\terms{Algorithms, Economics, Theory}

%Keywords are your own choice of terms you would like the paper to be indexed by.

%\keywords{Optimal mechanism design, multi-armed bandit mechanism, multi-unit procurement, strategic agents}

\section{Introduction}
\label{sec:intro}
\noindent Auction based mechanisms are widely used to allocate goods or services in the presence of strategic agents. In different contexts, the auctioneer may have different goals such as welfare maximization or utility maximization or revenue maximization or cost minimization. Auction theory generally assumes that the players are symmetric which means they are distinguished only by privately held types such as costs, valuations, or capacities. 
The theory does not consider the ``experience'' of an auctioneer resulting from the consumption of the commodity or service. The experience can be uncertain and not known upfront. 
For example, consider a hospital (auctioneer) interested in procuring a large number of units of a single generic drug from various pharmaceuticals who can supply limited quantities at different production costs. The quality of the procured generic drug from a supplier can depend on several parameters such as methodology used in preparation and other parameters which are inherent to the supplier. In this example and several other real world scenarios, there is an inherent heterogeneity amongst  services or items
procured from different agents.  Therefore, we can attribute to every agent an inherent quality which is a measure of the perceived experience or reward. Thus, in order to maximize her utility, the auctioneer needs to minimize her payments at the same time ensure a required quality of service.
If the qualities from different agents are observed repeatedly, the auctioneer can  learn the quality of the agents for future optimization.

A strong motivation for this work comes from the setting of crowdsourcing. The quality of human generated data or labels is an important input for an AI process or a machine learning system.  With the advent of several crowdsourcing marketplaces, such inputs are now obtained at much less cost from a global pool of heterogeneous crowd workers. These human workers have different quality levels and can be strategic about their costs. The risk of low quality levels is mitigated via learning algorithms which can predict high quality workers while strategic behavior of crowd workers can be addressed via mechanism design. Thus, the auctioneer here is a requester who seeks to procure tasks from strategic crowd workers with privately held costs, privately held capacities, and unknown qualities. 

Motivated by situations such as above, we consider a procurement scenario where a buyer (or auctioneer) wishes
to procure multiple units of a service or item from a pool of heterogeneous
agents with unknown qualities, privately held costs, and  privately held limited capacities. 
Our goal is to design a procurement auction that
learns the qualities of the agents, elicits true costs and capacities from the
agents,  and maximizes the expected utility of the auctioneer.  
If the agents are honest in reporting their costs
and capacities, the classical   Multi-Armed-Bandit (MAB) techniques can be used to learn the qualities. For example, Tran-Thanh et. al.~\cite{THANH14} have proposed a greedy approach to learn the qualities of the crowd workers. On the other hand, if all the agents have the same quality that is common knowledge but with strategic costs and capacities, the auctioneer can deploy the techniques available in the literature   \cite{IYENGAR08,SUJIT13} to elicit true costs and capacities. In the setting considered in this paper, in addition to strategic costs and capacities, we also address heterogeneity amongst agents and moreover we learn their qualities. %In the setting addressed in this paper, we are required to tackle not only unknown qualities but also strategic costs and capacities.

Learning in the presence of strategic agents in a multi armed bandit (MAB) setting leads to \emph{MAB mechanisms} \cite{BABAIOFF09}. In this paper, we take a detour from current MAB mechanism theory in two ways. (i) We propose an optimal MAB mechanism that performs nearly as well as an optimal auction with full information, whereas the current literature mainly focuses on social welfare maximization (ii) We provide a characterization for a weaker notion of truthfulness i.e. stochastic Bayesian incentive compatibility that can potentially achieve better regret bounds. More importantly, while the existing research is also limited to learning with agents having single dimensional private information,  we design an MAB mechanism when the agents' private information is two dimensional. In particular, following are the contributions of this paper:
\begin{itemize}[leftmargin=0.4cm]
\itemsep0em 
\item We first explore the case of heterogeneous agents with known qualities and provide a characterization for any Bayesian Incentive Compatible (BIC) and Individual Rational (IR) mechanism in a bidimensional setting. Using this characterization, we provide the footprint for a mechanism to be BIC, IR and maximizes the expected utility of the auctioneer (Theorem \ref{thm:offline_payment}). We then propose an optimal mechanism \CHOPT which is in fact dominant strategic incentive compatible (DSIC) and IR (Theorem \ref{thm:chopt_dsic}).
\item We next take up the  case when the qualities are unknown and  derive sufficient conditions for an allocation rule to be implemented in stochastic BIC and IR (Theorem \ref{lemma:online_payment}).\footnote{Note that, this is  sufficient condition and the complete characterization is still open.} This leads to a learning mechanism \CUCB that is stochastic BIC and IR (Theorem \ref{thm:chucb_bic}).
We evaluate \CUCB through simulations and show that the expected utility of an auctioneer adopting \CUCB mechanism approaches that of the omniscient \CHOPT.
\end{itemize}

\section{Positioning of our Work}
\label{sec:related_work}

\noindent An extensive study of auction theory and mechanism design can be found in \cite{KRISHNA09}. 
The notion of optimal auction was introduced by \citet{MYERSON81}. Subsequently, there were many significant results in single parameter domains, however, the multiple parameter domain was unexplored until recently. The readers are referred to \cite{mishra2012multidimensional,HARTLINE13} for more details on optimal multi-dimensional mechanism design. The settings addressed in most of the literature assume additive valuation. In our work, cost and capacity parameters constitute the private information and the valuation of the agents is not additive in these two parameters. Notably, \citet{IYENGAR08} have designed optimal single item multi unit auction for capacitated bidders and this is further developed by \citet{SUJIT13} for multi-item multi unit auctions. 
However, as pointed out in Section \ref{sec:intro}, the above works~\cite{IYENGAR08,SUJIT13}  assume that all agents are of the same quality. In our setting, the agents are heterogeneous and their qualities need to be learnt. 

If we assume honest agents, the multi-armed-bandit theory~\cite{LAI85,AUER00}
is applicable to learn the qualities of the agents. Upper confidence bound based algorithms have been designed to learn unknown quantities with logarithmic regrets \cite{BUBECK12}. In the specific context of crowdsourcing, much research has been carried out for learning qualities of the crowd workers~\cite{HO12,HO13,ABRAHAM13,LONG12,LONG13,HO14,SINGLA13,BADANADIYURU12,SINGER13,SHIPRA14}. In a pure learning setting devoid of strategic play, the closest setting to ours is the one in \citet{THANH14} which studies
the problem in the context of crowdsourcing to maximize the number of successful tasks under a fixed budget. Note that all the above papers assume costs are known.
% and many papers do not impose any restriction on how many times a worker can be hired. 

A learning algorithm can be potentially manipulated by a strategic agent so as to increase  utility. This problem is addressed using MAB mechanism design theory~\cite{BABAIOFF09,DEVANUR09,GATTI12,SUJIT12,SHWETA14,BABAIOFF10,DEBMALYA14,BABAIOFF13}. Most of the literature in this space (except \cite{BABAIOFF13}) considers strategic agents with single dimensional private information and seeks to maximize social welfare. Our work, on the other hand, seeks to maximize the expected utility of the auctioneer. The work in \cite{BABAIOFF13} considers a multi-parameter setting and seeks to maximize welfare, but with an additive valuation model where the valuation of each agent is a linear combination of different private values. Our work is different from \cite{BABAIOFF13} as we aim to design an optimal auction in a capacitated setting where additive valuations do not apply. 
%The main challenge in an MAB mechanism is payment computation. Classical mechanisms like VCG mechanisms cannot 
%directly be applied as the stochastic information about agents is unknown and therefore 
%externalities cannot be computed.  
%To overcome this, we adapt the tools developed in \cite{BABAIOFF10} based on a self sampling procedure for the setting of this paper.

%In this work, we design an optimal auction for heterogeneous capacitated agents with unknown qualities. 
%The strategic parameters are bidimensional and we consider a non-additive valuation settings. 
%

\section{Notation and Preliminaries}
\label{sec:not}
\noindent An auctioneer wishes to procure $L$ units of an item from an agent pool $N$ = $\{1,2,\ldots,n\}$. 
Let $q_i \in [0,1]$ represent the quality of agent $i$, let $c_i \in [\underline{c}_i,\overline{c}_i]$ be his true cost and let $k_i \in [\underline{k}_i,\overline{k}_i]$ represent the maximum number of units an agent can provide or his true capacity. Let, $q$, $c$, $k$ denote the vectors of qualities, costs and capacities respectively. We consider a linear reward function for the auctioneer and she obtains an expected reward of $Rq_i$ on procuring an unit from agent $i$ where $R$ is a fixed positive real number.

In this work, we make an important and reasonable assumption that the agent is not allowed to over-report his capacity. This is because if the auctioneer allocates the agent beyond his capacity, it is detected eventually when the agent fails to deliver. This could lead to imposition of a high penalty or may lead to blacklisting the agent from further participation.  In contrast to over-reporting, under-reporting of capacity cannot be detected. In the absence of proper incentives, an agent can create virtual scarcity of agents by under-reporting his capacity which can benefit him.

 We denote the reported cost by $\hat{c}_i \in [\underline{c}_i,\overline{c}_i]$ and the reported capacity by $\hat{k}_i \in [\underline{k}_i,k_i]$. Let $b_i = (\hat{c}_i,\hat{k}_i)$ denote the bid of agent $i$ and the bid vector of all the agents except $i$ is denoted by $b_{-i}$. The objective of the auctioneer is to maximize the expected reward from $L$ units of the item and at the same time also minimize the payments to the agents, ensuring that from each agent $i$ at most $\hat{k}_i$ units are procured. If all the parameters are known, then one can solve the following optimization problem which maximizes the utility of the auctioneer:
\begin{align}
\max \sum_{i=1}^n \bigg( x_iRq_i - t_i\bigg)\; \text{s.t.}\ x_i \in \{0,1, \ldots, \hat{k}_i\}\;\ \text{,}\ \sum_i x_i \le L, \label{eq:optdef}
 \end{align}
\noindent where, $x_i$ represents the number of units that are procured from an agent $i$ and $t_i$ denotes the payment given to an agent $i$. The total number of units procured from the agents $x = (x_1,x_2,\ldots,x_n)$ (allocation) and the payments made to the agents $t = (t_1,t_2,\ldots,t_n)$ form the mechanism denoted by $\mathcal{M} = (x,t)$. Note that the allocation $x$ and payment $t$ depend on the bids reported by the agents and the qualities. We assume an independent private value model, and that the joint probability density function denoted by $f_i(c_i,k_i)$ is common knowledge.
Let $X$ and $T$ denote the expected allocations and expected payments when expectation is taken over bids of other agents.
That is, $X_i(\hat{c}_i,\hat{k}_i;q_i)$ represents the expected number of units procured from agent $i$ when he bids cost per item  $\hat{c}_i$, bids capacity $\hat{k}_i$ and the quality is $q_i$. Similarly $T_i$'s are defined. 
We now define some desirable properties for a mechanism if qualities were known.

\noindent\begin{definition}{\bf{(Bayesian Incentive Compatible)}}
A mechanism is called \emph{ Bayesian Incentive Compatible} (BIC) if reporting truthfully gives an agent highest expected utility when the other agents are truthful, with the expectation taken over type profiles of other agents. Formally, $\forall i \in N, \forall \hat{c}_i, c_i \in [\underline{c}_i,\overline{c}_i], \forall \hat{k}_i \in [\underline{k}_i,k_i]$,
\begin{align}
\nonumber U_i(c_i,k_i, c_i, k_i;q) \geq U_i(\hat{c}_i, \hat{k}_i,c_i,k_i;q),
 \end{align}
 where, $U_i(\hat{c}_i,\hat{k}_i, c_i, k_i;q) = \mathbb{E}_{b_{-i}}[c_ix_i(\hat{c}_i,\hat{k}_i;q) + t_i(\hat{c}_i,\hat{k}_i;q)]$
\end{definition}
\noindent\begin{definition}{\bf{(Dominant Strategy Incentive Compatible)}}
 A mechanism is called \emph{Dominant Strategy Incentive Compatible} (DSIC) if reporting truthfully gives every agent highest utility irrespective of the bids of the other agents. Formally, $\forall i \in N, \forall \hat{c}_i, c_i \in [\underline{c}_i,\overline{c}_i], \forall \hat{k}_i \in [\underline{k}_i,k_i]$, $\forall \hat{c}_{-i},\ \forall \hat{k}_{-i}$,
  \begin{align*}
 u_i(c_i,\hat{c}_{-i},k_i,\hat{k}_{-i}, c, k;q) \geq u_i(\hat{c}_i,\hat{c}_{-i}, \hat{k}_i,\hat{k}_{-i},c,k;q)
 \end{align*}
 where $u_i(\hat{c}_i,\hat{c}_{-i},\hat{k}_i,\hat{k}_{-i}, c, k;q) = c_ix_i(\hat{c},\hat{k};q) + t_i(\hat{c},\hat{k};q)$ is the utility when the true bid profile is ${c,k}$ and agent $i$ reports ${\hat{c}_i,\hat{k}_i}.$
\end{definition}

\begin{definition}{\bf{(Individually Rational)}}
A mechanism is called \emph{Individually Rational} (IR) if no agent derives negative utility by participating in the mechanism. Formally, $\forall i \in N, \forall c_i \in [\underline{c}_i,\overline{c}_i], \forall k_i \in [\underline{k}_i,k_i]$,
\begin{align*}
u_i(c_i,k_i, c, k;q) \geq 0
\end{align*}
\end{definition}

\begin{definition}{\bf{(Optimal Mechanism)}}
A mechanism $\mathcal{M} = (x,t)$ is called optimal if it maximizes~\cref{eq:optdef} subject to BIC and IR.
\end{definition}

%We assume an independent private value model, where each agent derives his capacity and cost from a distribution which is independent of other agents. We also assume that the joint probability density function which we denote by $f_i(c_i,k_i)$ is common knowledge.

\section{Auction with Known Qualities}
\label{sec:offline}
\noindent We now derive the characterization for any mechanism to be BIC and IR when the qualities are known.
\subsection{Characterization}
\noindent In the setting considered in the paper, as described in~\cref{sec:not}, VCG mechanisms can be used to elicit the costs and capacities from the agents and it satisfies DSIC, IR. However, VCG mechanisms maximize social welfare and may or may not be utility maximizing for the auctioneer~\cite{rothkopf2007thirteen}.  

Any allocation should be compensated with at least the cost incurred by the agent, irrespective of the quality of the unit procured. We propose to pay a premium to each agent above his true cost so as to incentivize him to report costs and capacities truthfully. We define $\forall i \in N,$
\begin{align*}
\rho_i(b_i;q)=T_i(b_i;q)-\hat{c}_iX_i(b_i;q), \mbox{ where } 
		b_i=(\hat{c}_i,\hat{k}_i).
\end{align*}
The utility of an agent $i$ with bid $b_i$ is given as,
\begin{align}
U_i(b_i,c_i,k_i;q) &= T_i(b_i;q) - c_i X_i(b_i;q) \nonumber\\
		&= \rho_{i}(b_i;q) -(c_i-\hat{c}_i)X_i(b_i;q) \label{eq:rho_utility}
\end{align}
Thus $\rho_i$ represents the offered utility when all the agents are truthful. With the above offered incentive, we have the following theorem. 
\begin{theorem}
\label{thm:bic_ir}
A mechanism is BIC and IR iff $\forall i \in N$,
	\begin{enumerate}[leftmargin=0.4cm]
\itemsep0em 
		\item  	{$X_i(\hat{c}_i,\hat{k}_i;q)$ is non-increasing in $\hat{c}_i,\ \forall q \mbox{ and } \forall \hat{k}_i \in [\underline{k}_i,k_i]$}\label{thm:mon-cond2}.        
		\item {$\rho_{i}(\hat{c}_i,\hat{k}_i;q)$ is non-negative, and non-decreasing in $\hat{k}_i$ $\forall\;q $ and $\forall\;\hat{c}_i\;\in\;[\underline{c}_i,\bar{c}_i]$}\label{thm:mon-cond1}		
		\item {$\rho_{i}(\hat{c}_i,\hat{k}_i;q) = \rho_{i}(\bar{c}_i,\hat{k}_i;q) + \int_{\hat{c}_i}^{\overline{c}_i}X_i(z,\hat{k}_i;q)dz $} \label{thm:utl-form}	
	\end{enumerate}
\end{theorem}
We refer to the above three statements as conditions \ref{thm:mon-cond2}, \ref{thm:mon-cond1} and \ref{thm:utl-form} respectively. \\

{\em \noindent Proof:}
\noindent To prove the necessity part, we first observe due to BIC, 
\begin{align*}
&U_i(\hat{c}_i,\hat{k}_i,c_i,k_i;q) \leq  U_i(c_i,k_i,c_i,k_i;q) \qquad\forall(\hat{c}_i,\hat{k}_i) \mbox{ and }(c_i,k_i)\\
&\implies U_i(\hat{c}_i,k_i,c_i,k_i;q)\leq U_i(c_i,k_i,c_i,k_i;q)
\end{align*}
We assume $\hat{c}_i>c_i.$ The proof follows in identical lines otherwise. From \cref{eq:rho_utility},
\begin{align*}
U_i(\hat{c}_i,k_i,c_i,k_i;q) = U_i(\hat{c}_i,k_i,\hat{c}_i,k_i;q) 
+ (\hat{c}_i-c_i)X_i(\hat{c}_i,k_i;q),
\end{align*}
which implies that,
\begin{align*}
\frac{U_i(\hat{c}_i,k_i,\hat{c}_i,k_i;q)-U_i(c_i,k_i,c_i,k_i;q)}{\hat{c}_i-c_i}
 \leq -X_i(\hat{c}_i,k_i;q).
\end{align*}

Similarly using $U_i(c_i,k_i,\hat{c}_i,k_i;q) \leq U_i(\hat{c}_i,k_i,\hat{c}_i,k_i;q)$,
\begin{align}
-X_i(c_i,k_i;q) &\leq\frac{U_i(\hat{c}_i,k_i,\hat{c}_i,k_i;q)-U_i(c_i,k_i,c_i,k_i;q)}{\hat{c}_i-c_i}\nonumber \\
&\leq-X_i(\hat{c}_i,k_i;q).\label{eq:mono1}
\end{align}
\noindent Taking limit $\hat{c}_i\rightarrow c_i,$ we get,
\begin{align}
\frac{\partial U_i(c_i,k_i,c_i,k_i;q)}{\partial{c}_i} = -X_i(c_i,k_i;q). 
\label{eq:pde}
\end{align}
Equation (\ref{eq:mono1}) implies, $X_i(c_i,k_i;q)$ is 
non-increasing in $c_i$.  This proves condition \ref{thm:mon-cond2} of the 
theorem in the forward direction. When the worker bids truthfully, 
from Equation (\ref{eq:rho_utility}),
\begin{align}
\rho_{i}(c_i,k_i;q)=U_i(c_i,k_i,c_i,k_i;q).\label{eq:rho1}
\end{align}
For BIC, Equation (\ref{eq:pde}) should be true. So,
\begin{align}
\rho_{i}(c_i,k_i;q)=\rho_{i}(\bar{c}_i,k_i;q)+\int_{c_i}^{\bar{c}_i}X_i(z,k_i;q)dz\label{eq:rho2}
\end{align}
This proves condition \ref{thm:utl-form} of the theorem. BIC also requires,
\begin{align*}
k_i \in \argmax_{\hat{k}_i\in[\underline{k}_i,k_i]} 
U_i(c_i,\hat{k}_i,c_i,k_i;q)
\;\forall\; c_i\;\in\;[\underline{c}_i,\bar{c}_i]
\end{align*}

\noindent This implies, $\forall c_i,\;\rho_{i}(c_i,k_i;q)$ should be 
non-decreasing in $k_i$. The IR conditions (Equation(\ref{eq:rho1})) imply $$\rho_{i}(c_i,k_i;q)\geq 0.$$
This proves condition \ref{thm:mon-cond1} of the theorem. Thus, these three 
conditions are necessary for BIC and IR properties.
%%%%%%%%%%%%% Sufficient part of the proof
We now prove the sufficiency. Consider
\begin{align*}
U_i(c_i,k_i,c_i,k_i;q)=\rho_i(c_i,k_i;q) \geq 0.
\end{align*}
So the IR property is satisfied. We assume $\hat{c}_i>c_i.$ The proof is similar for the case $\hat{c}_i<c_i.$ To establish BIC, consider:
\begin{align*}
&U_i(\hat{c}_i,\hat{k}_i,c_{i},k_i;q) \\
&=\rho_{i}(\hat{c}_i,\hat{k}_i;q)+(\hat{c}_i-c_i)X_i(\hat{c}_i,\hat{k}_i;q)\tag*{(By Defn)}\nonumber \\
&= \rho_{i}(\bar{c}_i,\hat{k}_i;q) 
		+ \int_{\hat{c}_i}^{\bar{c}_i}X_i(z,\hat{k}_i;q)dz 
		+ (\hat{c}_i-c_i)X_i(\hat{c}_i,\hat{k}_i) \tag*{(By hypothesis)} \nonumber \\
&= \rho_{i}(\bar{c}_i,\hat{k}_i;q) 
		+ \int_{c_i}^{\bar{c}_i}X_i(z,\hat{k}_i;q)dz \\
& \qquad \qquad - \int_{c_i}^{\hat{c}_i}X_i(z,\hat{k}_i;q)dz 
		+ (\hat{c}_i-c_i)X_i(\hat{c}_i,\hat{k}_i;q)\nonumber \\
&\leq \rho_{i}(c_i,\hat{k}_i;q)
		\tag*{($X_i$ is non-increasing in $c_i$)}
	\nonumber \\
&\leq \rho_{i}(c_i,k_i;q) \tag*{( as $\rho_{i}$ is non-decreasing in $k_i$)} \nonumber \\
&= U_i(c_{i},k_i,c_i,k_i;q) 		 \nonumber
&\hfill \blacksquare
\end{align*}
\subsection{Sufficiency Conditions for Optimality}
\noindent We now present sufficiency conditions for an IR, BIC mechanism to be optimal.  Let $F_i(c_i|k_i)$ and $f_i(c_i|k_i)$ denote respectively the cumulative distribution and probability density function of cost of an agent $i$ given the capacity. 
%We use $\bar{c}$ for the vector $(\bar{c}_1,\dots, \bar{c}_n )$ and $\underline{c}$ for the vector $(\underline{c}_1,\dots, \underline{c}_n )$. We also use $\bar{k}$ and $\underline{k}$ which are defined analogously.
\begin{theorem}
\label{thm:offline_payment}
Suppose the allocation rule maximizes
\begin{align}
&\sum_{i=1}^{n} \int_{\underline{c}_1}^{\bar{c}_1}\ldots \int_{\underline{c}_n}^{\bar{c}_n}\int_{\underline{k}_1}^{\bar{k}_1} \ldots\int_{\underline{k}_n}^{\bar{k}_n} \Bigg( Rq_i - \bigg(c_i + \frac{F_i(c_i|k_i)}{f_i(c_i|k_i)} \bigg) \Bigg) \nonumber \\
&x_i(c_i,k_i,c_{-i},k_{-i}) f_1(c_1,k_1) \ldots f_n(c_n,k_n) \,dc_1\ldots dc_n \, dk_1 \ldots dk_n  \label{opt_stmt}
\end{align}
\normalsize
\noindent subject to conditions \ref{thm:mon-cond2} and \ref{thm:mon-cond1} of \Cref{thm:bic_ir}. Also suppose that the payment is given by
\begin{align}
T_i(c_i,k_i;q) = c_iX_i(c_i,k_i;q) + \int_{c_i}^{\overline{c}_i}
	X_i(z,k_i;q)dz \label{eqn:opt_payment}
\end{align}
\noindent then such a payment scheme and allocation scheme constitute an optimal auction satisfying BIC and IR.
\end{theorem}

\begin{pfof}
The auctioneer's objective is to maximize her expected utility which is:
\begin{align}
&\sum_{i=1}^{n}\int_{\underline{c}_1}^{\bar{c}_1}\ldots \int_{\underline{c}_n}^{\bar{c}_n}\int_{\underline{k}_1}^{\bar{k}_1} \ldots\int_{\underline{k}_n}^{\bar{k}_n} \big[ R q_i x_i(b;q) -t_i(b;q)\big] \nonumber \\
& f_1(c_1,k_1)\ldots f_n(c_n,k_n) dc_1\ldots dc_n \, dk_1 \ldots dk_n  \nonumber \\
&=\sum_{i=1}^{n}\int_{\underline{c}_1}^{\bar{c}_1}\ldots \int_{\underline{c}_n}^{\bar{c}_n}\int_{\underline{k}_1}^{\bar{k}_1} \ldots\int_{\underline{k}_n}^{\bar{k}_n} \big[x_i(b;q) (Rq_i  -c_i + c_i)-t_i(b;q)\big]  \nonumber \\
&\qquad f_1(c_1,k_1)\ldots f_n(c_n,k_n) dc_1\ldots dc_n \, dk_1 \ldots dk_n  \nonumber \\
&=\sum_{i=1}^{n}\int_{\underline{c}_1}^{\bar{c}_1}\ldots \int_{\underline{c}_n}^{\bar{c}_n}\int_{\underline{k}_1}^{\bar{k}_1} \ldots\int_{\underline{k}_n}^{\bar{k}_n} \big( c_i x_i(b;q)-t_i(b;q) \big)\nonumber \\
& f_1(c_1,k_1)\ldots f_n(c_n,k_n) dc_1\ldots dc_n \, dk_1 \ldots dk_n  \nonumber \\
&+\sum_{i=1}^{n} \int_{\underline{c}_1}^{\bar{c}_1}\ldots \int_{\underline{c}_n}^{\bar{c}_n}\int_{\underline{k}_1}^{\bar{k}_1} \ldots\int_{\underline{k}_n}^{\bar{k}_n} \Bigg( Rq_i - c_i  \Bigg)x_i(c_i,k_i,c_{-i},k_{-i}) \nonumber \\
&\qquad  f_1(c_1,k_1) \ldots f_n(c_n,k_n) \,dc_1\ldots dc_n \, dk_1 \ldots dk_n  \label{opt_stmtint}
\end{align}
The second term of \cref{opt_stmtint} is already similar to the desired form of the objective function of auctioneer given in \cref{opt_stmt}. We now use conditions \ref{thm:mon-cond2} and \ref{thm:utl-form} of \Cref{thm:bic_ir} to arrive at the result. Consider the first term,
{\allowdisplaybreaks
\begin{align}
&\int_{\underline{c}_1}^{\bar{c}_1}\ldots \int_{\underline{c}_n}^{\bar{c}_n}\int_{\underline{k}_1}^{\bar{k}_1} \ldots\int_{\underline{k}_n}^{\bar{k}_n} \big( c_i x_i(b;q)-t_i(b;q) \big) \nonumber \\ &f_1(c_1,k_1)\ldots f_n(c_n,k_n) dc_1\ldots dc_n \, dk_1 \ldots dk_n  \nonumber \\
&= - \int_{\underline{k}_i}^{\bar{k}_i}\int_{\underline{c}_i}^{\bar{c}_i} \rho_i(c_i,k_i;q) f_i(c_i,q_i) dc_i \, dk_i \tag*{(Integrating out $b_{-i}$)} \nonumber \\
&=  -\int_{\underline{k}_i}^{\bar{k}_i}\int_{\underline{c}_i}^{\bar{c}_i} \bigg(\rho_i(\bar{c}_i, k_i)  +  \int_{c_i}^{\bar{c}_i} X(z,k_i;q) dz\bigg) \, f_i(c_i,k_i)  dc_i \, dk_i \tag*{(As we need truthfulness)} \\
&=  -\int_{\underline{k}_i}^{\bar{k}_i}\int_{\underline{c}_i}^{\bar{c}_i} \rho_i(\bar{c}_i, k_i) f_i(c_i,k_i)  dc_i \, dk_i \nonumber \\ 
& \qquad - \int_{\underline{k}_i}^{\bar{k}_i}\int_{\underline{c}_i}^{\bar{c}_i} 
 X_i(z,k_i;q) dz \int_{\underline{c}_i}^{z}  \, f_i(c_i|k_i)  dc_i \; f_i(k_i) dk_i \tag*{(Changing order of integration)}\nonumber \\  
 &=  -\int_{\underline{k}_i}^{\bar{k}_i}\int_{\underline{c}_i}^{\bar{c}_i} \rho_i(\bar{c}_i, k_i) f_i(c_i,k_i)  dc_i \, dk_i \nonumber \\ 
& \qquad - \int_{\underline{k}_i}^{\bar{k}_i}\int_{\underline{c}_i}^{\bar{c}_i} 
 X_i(z,k_i;q) F_i(z|k_i) dz  f_i(k_i) dk_i \nonumber \\  
 &=  -\int_{\underline{k}_i}^{\bar{k}_i}\int_{\underline{c}_i}^{\bar{c}_i} \rho_i(\bar{c}_i, k_i) f_i(c_i,k_i)  dc_i \, dk_i \nonumber \\ 
& \qquad - \int_{\underline{k}_i}^{\bar{k}_i}\int_{\underline{c}_i}^{\bar{c}_i} 
 X_i(c_i,k_i;q) \frac{F_i(c_i|k_i)}{f_i(c_i|k_i)}   f_i(c_i, k_i) dc_i \, dk_i \label{eq-inter}
\end{align}
}
\noindent The last step is obtained by relabeling the variable of integration and simplifying.

Here, $\rho_i(\bar{c}_i, k_i)$ denotes the utility of an agent $i$ when his true type is $(\bar{c}_i, k_i)$. With this type profile, the auctioneer by paying $\bar{c}_i$ can ensure both IR and IC, hence we can set $\rho_i(\bar{c}_i, k_i) = 0, \forall k_i \in [\underline{k}_i,\bar{k}_i]$. Applying this in the above equation, we get that the objective function of the auctioneer is similar in form to ~\cref{opt_stmt}. Consider Condition \ref{thm:utl-form} of \Cref{thm:bic_ir}, and set $\rho_i(\bar{c}_i, k_i) = 0$, we get \cref{eqn:opt_payment}.
By construction, the mechanism is BIC and IR. And, since the auctioneer's expected utility is maximized the mechanism is optimal.
\end{pfof}

Analogous to the literature on optimal auction~\cite{SUJIT13,IYENGAR08,MYERSON81}, we assume regularity on our type distribution as follows.
%\begin{definition}[Regularity:]
%If for all $i= 1,2,\dots, n$ , we assume that the virtual cost function as given by
%{\small
%$$ H_i(c_i,k_i) := c_i + \frac{F_i(c_i|k_i)}{f_i(c_i|k_i)}$$}
%is non-decreasing in $c_i$ and non-increasing in $q_i$, we say regularity assumption holds true.
%\end{definition}

%\sg{how about like this}
\begin{definition}[Regularity]
We define  the virtual cost function $\forall i \in N$ as
\begin{align*}
H_i(c_i,k_i) := c_i + \frac{F_i(c_i|k_i)}{f_i(c_i|k_i)}
\end{align*}
We say that a type distribution is regular if $\forall i$, $H_i$ is non-decreasing in $c_i$ and non-increasing in $k_i$.
\end{definition}
This assumption is not restrictive in single dimension setting as standard techniques of ironing are available~\cite{MYERSON81}. The ironing techniques can also be applied in bidimensional setting whenever the marginal cost distribution is independent of marginal capacity distribution.

\subsection{\CHOPT: An Optimal Auction}
\noindent We now present our mechanism \CHOPT give in \Cref{alg:offline_mechanism}. 

\begin{algorithm}[h]{}
\SetKwBlock{Begin}{}{}
\small
\SetAlgoLined
\DontPrintSemicolon
\caption{\CHOPT Mechanism}\label{alg:offline_mechanism}
\KwIn {$\forall i$, Bids $b_i = (\hat{c}_i \, \hat{k}_i)$, reward parameter $R$}
\KwOut {An optimal, DSIC, IR Mechanism $\mathcal{M} = (x,t)$}
Allocation is given by $x$ = ALLOC($N,\hat{c}, \hat{k}, q, L$)\;
\For{$ i \in N$ $\&\&$ $x_i\neq 0$}
{
$G_i := Rq_i - H_i(b_i)$\;
$y=$ ALLOC($N\setminus\{i\},\hat{c}_{-i}, (\hat{k}_{-i}-x_{-i}), q_{-i}, x_i$) \label{allocminusi}\;
Payment to i, $t_i$ = $\displaystyle \sum_{k \in N\setminus\{i\}} y_k \max(G_i^{-1}(Rq_k - H_k(b_k)),\bar{c}_i) + \big(x_i- \sum_k y_k\big) \bar{c}_i $\;
}

\setcounter{AlgoLine}{0}
\hrule
\vspace{0.1cm}
Subroutine: ALLOC($N^\tau, c^\tau,k^\tau, q^\tau, L^\tau$) \nonumber \;
\hrule
\vspace{0.1cm}
\KwIn {
 $\langle N^\tau, c^\tau,k^\tau, q^\tau, L^\tau \rangle$ where\\
$N^\tau$ =: Set of agents,\\
$c^\tau$ =: Bid vector of costs, \\
$k^\tau$ =: Bid vector of capacities, \\
$q^\tau$ =: Vector of qualities, \\
$L^\tau$ =: Total number of units being allocated. \\
}
\KwOut {Vector $x$ of units allocated to each agent.}
\For{$ \kappa \in N^\tau$}
{
$H_\kappa(c^\tau_\kappa, k^\tau_\kappa) = c^\tau_\kappa + \frac{F_\kappa(c^\tau_\kappa|k^\tau_\kappa)}{f_\kappa(c^\tau_\kappa|k_\kappa)}$\;
$G_\kappa := Rq^\tau_\kappa - H_\kappa(c^\tau_\kappa, k^\tau_\kappa)$\;
}
$(a_1,a_2,\ldots)$ = Sorted indices of agents in $N^\tau$ in non-increasing order of $G_\kappa$\;
$x=0$\;
$L^{(1)}=L^\tau$\;
\For{$ 1 \leq \eta \leq |N^\tau|$ $\&\&$  $G_{a_\eta} \geq 0$\label{ln:zero}}
{
$x_{a_\eta} = \max (k^\tau_{a_\eta},L^{(\eta)} )$\;
$L^{(\eta+1)}= L^{(\eta)}-x_{a_\eta}$\;
}
\end{algorithm}
\begin{theorem}
Mechanism \CHOPT is optimal, DSIC and IR. 
\label{thm:chopt_dsic}
\end{theorem}
{\em \noindent Proof: } We will prove that \CHOPT satisfies~\Cref{thm:offline_payment}, which proves optimality, IR, and BIC. The allocation function (ALLOC) allocates maximum possible units to agents in decreasing order of $G$'s, which in turn maximizes \cref{opt_stmt}. This is because \cref{opt_stmt} is a linear combination of $G$'s. The monotonicity constraint  \ref{thm:mon-cond2} of \Cref{thm:bic_ir} is satisfied due to regularity.

Fix an agent $i$ with non-zero allocation. We will show that the payment given to the agent $i$ given by \CHOPT is the same as in~\cref{eqn:opt_payment}. We fix a bid profile $b_{-i}$, that yields non-zero allocation to agent $i$. The payment to agent $i$ for bid profile $(b_i, b_{-i})$ as per~\cref{eqn:opt_payment} is as follows. 
\begin{align}
t_i(c_i,k_i,b_{-i};q) = c_ix_i(c_i,k_i,b_{-i};q) + \int_{c_i}^{\overline{c}_i}
	x_i(z,k_i,b_{-i};q)dz
\label{eq:condpayment}
\end{align}
If expectation is taken on $b_{-i}$ for~\cref{eq:condpayment}, we get~\cref{eqn:opt_payment}. The interchange of integral and expectation required therein is valid due to Fubini's Theorem~\cite{royden1988real} as the integrand is finite and non-negative. We will show that \CHOPT computes this payment for any $b_{-i}$.

\begin{figure}[h!]
\centering
 \includegraphics[height=2.1in]{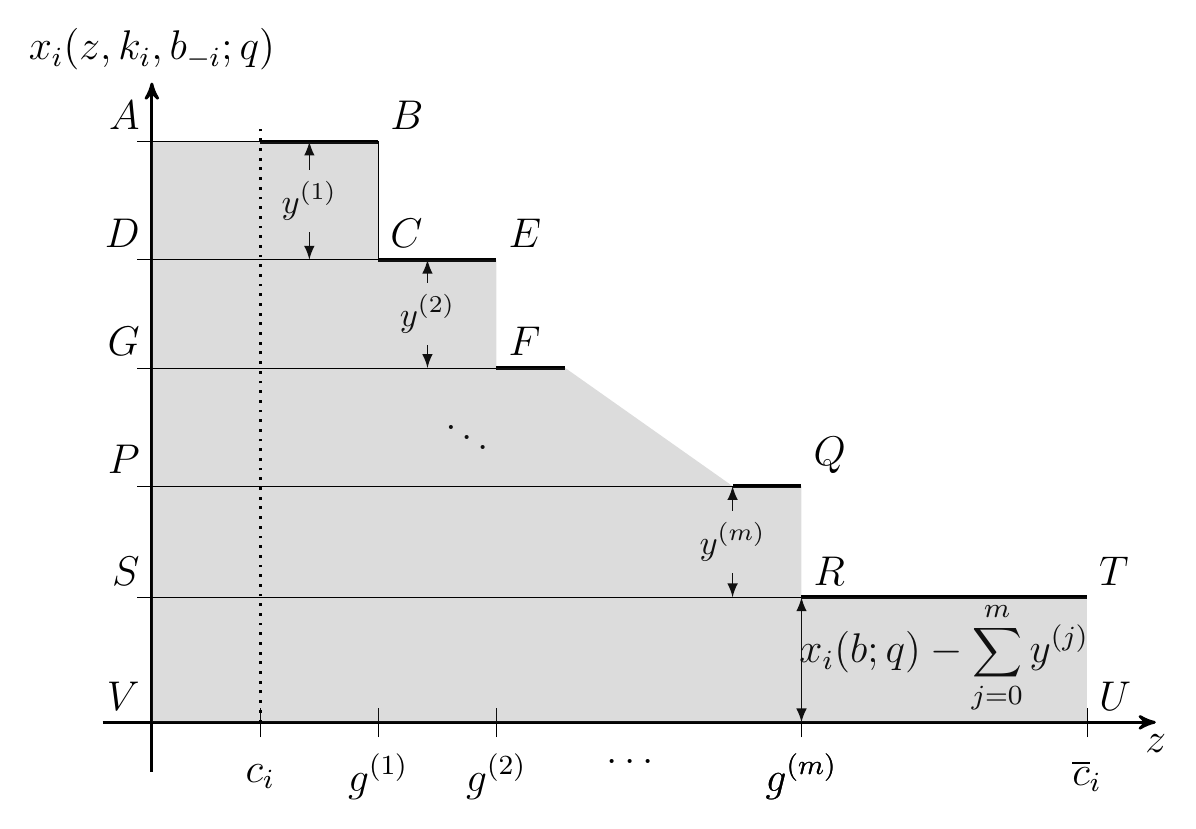}
\caption{Allocation to agent $i$ as function of his bid $z$}
 \label{fig:pfint}
\end{figure}

To compute RHS of~\cref{eq:condpayment}, we  first observe that when bidder $i$ alone increases his bid, he can lose some (or all) of the units allocated to him to bidders with lower values of $G$. Hence, the allocation to agent $i$ as a function of his bid $z \in [c_i, \bar{c}_i]$ is a step function as shown in Figure~\ref{fig:pfint}. And, the payment to be given to agent $i$ as per  ~\cref{eq:condpayment} is the shaded area.

Let $g^{(1)}<g^{(2)}<\ldots ... <g^{(m)}$ where $g^{(1)} >c_i$, $g^{(m)}< \overline{c}_i$, be the costs at which agent $i$ loses some more of his units. At these points, the allocation also dictates that an allocated agent $r$ either completely exhausts the units $x_i$ allocated previously to $i$ or he himself has no more capacity left.  

On the other hand, the payment scheme of \CHOPT first determines the allocation  of $x_i(c_i, k_i, c_{-i}, k_{-i})$ units in the absence of $i$ as given by \cref{allocminusi} of~\cref{alg:offline_mechanism}.

 Let $U =: \{j \in N \setminus \{i\}: y_j \neq 0\}$ where $y$ is the allocation to the worker set $N\setminus\{i\}$. We will partition the set $U$ into $V  =: \{j \in N \setminus \{i\}: y_j \neq 0 \mbox{, } G_i(\bar{c}_i)  < G_j < G_i(c_i) \}$ and $W  =: \{j \in N \setminus \{i\}: y_j \neq 0 \mbox{, } 0< G_j <G_i(\bar{c}_i)  \}$. With out loss of generality, we will assume $G_i(\bar{c}_i) \geq 0$, otherwise we will relabel $G_i^{-1}(0)$ as $\bar{c}_i$. No allocations are made to agents with negative value of $G$(see line~\ref{ln:zero} of ALLOC). Also, as allocation of $x_i$ units consider residual capacity $(\hat{k}_{-i}-x_{-i})$ (see line~\ref{allocminusi} of \cref{alg:offline_mechanism}), no agent with $G$ higher than $G_i(c_i)$ will have any capacity left. 

For the sake of simpler exposition, we will assume $U=V\cup W$, the proof follows similar lines otherwise. Let $(a_1,a_2, ... , a_m)$ as the indices of agents in $V$ sorted in non-increasing order of $G$. Now, agents are allocated units from $x_i$ in the order given by $(a_k)_{k=1}^{m}$. Now, it follows that $G_i^{-1}(Rq_{a_1} - H_{a_1}(b_{a_1}))=g^{(1)}$ and the allocation to this agent $a_1$ corresponds to $y^{(1)}$. This forms the term $y_{a_1}G^{-1}((Rq_{a_1} - H_{a_1}(b_{a_1}))$ of the payment to $i$ and corresponds to the area of rectangle $ABCD$. Similarly, the payment to $i$ due to $a_2$ corresponds area of rectangle $DEFG$. This holds for all agents in the set $V$ and rectangle $PQRS$ denotes the payment due to $a_m$. Finally, rectangle $STUV$ corresponds to agents in $W$ or units that are unallocated as there is no capacity left in the remaining agents. The latter is captured by the term $(x_i- \sum_k y_k)\bar{c}_i$. Hence proposed payment computes \cref{eqn:opt_payment} as we have shown it for any fixed $b_{-i}$. 

The offered utility $\rho_i$  when all agents are truthful is non-decreasing  in the true capacity $k_i$. This is due to the greedy nature of the allocation in ALLOC. Thus, condition \ref{thm:mon-cond1} of \Cref{thm:bic_ir} is satisfied. 

Thus, \CHOPT satisfies the \Cref{thm:offline_payment}. We therefore have that the proposed mechanism is BIC, IR, and optimal.

In respect of proving DSIC, we omit a formal proof due to space constraint and provide only a sketch. We note that the allocation is deterministic and the payment to agent $i$ does not depend on his bid directly and only depends via the allocation. Furthermore, the payments are computed based on the allocations that are made in the absence of $i$ for the $x_i$ units he has been allocated currently. For every unit, the agent is paid the best possible price he could have bid and still won the unit.  \hfill $\blacksquare$
\section{Auction with Unknown Qualities}
\label{sec:online}
\noindent This section addresses the problem when qualities are not known and are to be learnt. In order to maximize her utility, the auctioneer will procure units from agents in a sequential manner so that she can make future decisions based on the past learning history. We now discuss definitions relevant in this setting.
%\sj{heading of this subsection should be changed as we are not providing full characterizations}
%\sg{Just wondering do you want to call it a quality realization????or you prefer to
%be constent with literature?}
\begin{definition}[Reward Realization]
A reward realization $s$ is an $n\times L$ table where the $(i,j)$ entry represents an independent realization drawn from the true quality of $i^{th}$ agent when procuring the $j^{th}$ unit from him.
\end{definition}

Note that $(i,j)$ entry in reward realization indicates the quality of $i^{th}$ agent when $j^{th}$ unit is procured from him and not the $j^{th}$ unit procured by the requester. 

\begin{definition}[Stochastic BIC Mechanism]
We say that a mechanism $\mathcal{M}=(x,t)$ is Stochastic BIC if truth telling by any agent $i$ results in highest expected utility when expectation is taken over reward realizations and type profiles of other agents. Formally, $\forall \hat{c}_i \in [\underline{c}_i,\overline{c}_i], \hat{k}_i \in[\underline{k}_i,k_i],$
\begin{align*}
\mathbb{E}_{s}[U_i(c_i,k_i,c_i,k_i;s)] \ge \mathbb{E}_{s}[U_i(\hat{c}_i,\hat{k}_i,c_i,k_i;s)].
\end{align*}
\end{definition}

%where inner expectation is taken over ${(c_{-i},k_{-i})}$

\subsection{Sufficiency Conditions for Stochastic BIC}
\noindent We now provide sufficiency conditions for a mechanism to be stochastic BIC and IR. We begin by stating the modified characterization theorem for the learning setting. 
\begin{theorem}
\label{thm:online_bic}
Any mechanism that satisfies the following conditions $\forall i \in N,\ \forall s \in [0,1]^{n\times L}$, is stochastic BIC and IR. 
\begin{enumerate}[leftmargin=0.4cm]
\itemsep0em 
\item $X_i(c_i,k_i;s)$ is non-increasing in $c_i$, $\forall s \mbox{ and } \forall k_i \in [\underline{k}_i,k_i]$.
\item $\rho_{i}(\hat{c}_i,\hat{k}_i;s)$ non-negative, and non-decreasing in $\hat{k}_i\; \forall s$  and $\forall \hat{c}_i$ $\in [\underline{c}_i,\bar{c}_i].$	
\item $\rho_{i}(\hat{c}_i,\hat{k}_i;s) = \rho_{i}(\bar{c_i,}\hat{k}_i;s) + \int_{\hat{c}_i}^{\bar{c}_i}X_i(z,\hat{k}_i;s)dz$\label{online_payment}	
\end{enumerate}
\end{theorem}

The proof of the above theorem is similar to that of Theorem \ref{thm:bic_ir}. 
Instead of fixing a quality, we are now fixing a reward realization. The mechanism also remains stochastic BIC and IR when it satisfies~\Cref{thm:online_bic} and expectation is taken over reward realization.

We now discuss a set of natural properties which a mechanism in this space ideally have. It also turns out that these properties are sufficient to ensure BIC and IR.

\begin{definition}[\PRIN Allocation Rule]
An allocation rule $x$ is called a \PRIN Allocation if:
\begin{enumerate}[leftmargin=0.4cm]
\itemsep0em 
\item Allocation to any agent $i$ for the unit being allocated in round $j$, $x_i^j$, for any reward realization $s$ depends only on the agent's bids and the reward realization of $j$ units that are procured by the auctioneer so far and is non decreasing in terms of costs.
\item For the unit being allocated in round $j$ and for any three distinct agents $\{\alpha,\beta,\gamma \}$ such that $j^{th}$ round unit is allocated to $\beta$. A change of bid by agent $\alpha$  should not transfer allocation of $j^{th}$ round unit from $\beta$ to $\gamma$ if other quantities are fixed till $j$ units.
\item For all reward realizations $s$, $x_i(c_i,k_i;s)$ is non-decreasing with increase in capacity $k_i$ 
\end{enumerate}
\end{definition}

As mentioned earlier, these properties are natural. Property $1$ states that the allocation should not depend on any future success realizations which are not observed. Property $2$ is similar to Independent of Irrelevant Alternatives (IIA) property in the mechanism design theory i.e. if an agent $i$ changes his bid then it should not affect the allocations of other agents. Property 3 states the allocation rule doesn't penalize an agent with higher capacity, when other parameters are identical.

\begin{lemma}
\label{lemma:online_monotone}
If an allocation rule $x$ is \MakeLowercase{\PRIN} then, $\forall s$, and $\forall \hat{k}_i \in [\underline{k}_i,k_i]$, $x_i(c_i,\hat{k}_i;s)$ is non-increasing in $c_i$.
\end{lemma}
%The intuition is when a reward realization is fixed, the allocation can only increase by reducing the cost as the allocation rule is monotone in terms of cost when all the other quantities are fixed by the definition of \MakeLowercase{\PRIN}rule.\\

\begin{pfof}
By slight abuse of notation, let $x_i(c_i,t)$ denote the number of items procured by an agent $i$ with bid $c_i$ until $j$ items are procured. We need to prove that, 
\begin{align*}
x_i(c_i,j) \le x_i(c_i^-,j)\ \forall c_i^- \le c_i
\end{align*}
We will prove this by induction. At $j=1$, the condition trivially holds by the monotonicity property of \MakeLowercase{\PRIN} allocation rule. Thus, by induction hypothesis, $x_i(c_i,j) \le x_i(c_i^-,j)$ and we need to prove that $x_i(c_i,j+1) \le x_i(c_i^-,j+1)$. Without loss of generality, we will consider, $x_i(c_i,j) = x_i(c_i^-,j)$, otherwise the condition is trivially satisfied. 

In this case, we will show that $x_m(c_i,j) = x_m(c_i^-,j)\ \forall m$. Note that $x_m$ depends on bids of all the agents. Since the cost of other agents and capacities of all the agents are held fixed, we have dropped these dependence for notational convenience. Let $x_*(c_i,j)$ denote the number of units that are not procured by an agent $i$ until $j$ units, i.e. $x_*(c_i,j) = j - x_i(c_i,j)$, we will prove that for any two units $j$,$j'$:
\begin{align*}
x_*(c_i,j) = x_*(c_i^-,j') \implies x_m(c_i,j) = x_m(c_i^-,j')\ \forall m\ne i
\end{align*}
We prove the above statement using induction again. If $x_*(c_i,j)$ $= x_*(c_i,j') = 0$, that means all the items are procured by the agent $i$, the statement is clearly true. Thus, by induction hypothesis, $x_*(c_i,j) = x_*(c_i,j') = x_*$, then $x_m(c_i,j) = x_m(c_i^-,j')\ \forall m\ne i$. Now, suppose $x_*(c_i,j) = x_*(c_i^-,j') = x_* + 1$. Again by induction hypothesis, there exist latest rounds, $j_1 < j$ and $j'_1 < j'$ such that $\forall m'\ne i$ 
\begin{align*}
x_*(c_i,j_1) = x_*(c_i^-,j_1') = x_* \implies x_{m'}(c_i,j_1) = x_{m'}(c_i^-,j'_1)
\end{align*}
Since $j_1$ and $j'_1$ are the latest such rounds, units from $j_1+2$ to $j$ and $j'_1+2$ to $j'$ are procured only by agent $i$, thus we need to prove that allocation at round $j_1+1$ and $j'_1+1$ is same with bid $c_i$ and $c_i^-$ respectively. Since agent $i$ is not allocated at these rounds, by property $2$ of \MakeLowercase{\PRIN} allocation rule, the condition is satisfied. Thus, we have $x_i(c_i,j) = x_i(c_i^-,j) \implies x_*(c_i,j) = x_*(c_i^-,j) \implies x_m(c_i,j) = x_m(c_i^-,j)\ \forall m$

Since the reward realization is fixed, if number of allocations to all the agents is same till $j^{th}$ unit procured then by property $1$ of \MakeLowercase{\PRIN} allocation rule, we have $x_i(c_i,j+1) \le x_i(c_i^-,j+1)$. 
\end{pfof}

%\end{proof}
The following theorem guarantees a transformation of any \MakeLowercase{\PRIN} allocation rule into a stochastic BIC and IR mechanism.
\begin{theorem}
\label{lemma:online_payment}
For a \MakeLowercase{\PRIN} allocation rule, there exists a transformation that produces the transformed allocation ($\tilde{x}$) and payment ($\tilde{t}$) such that the resulting mechanism $\mathcal{M} = (\tilde{x},\tilde{t})$ is stochastic BIC and IR.
\end{theorem}

If we implement the following payment rule then we will get stochastic BIC by Theorem \ref{thm:online_bic}:
\begin{align}
\label{eq:truthfulness}
T_i(\hat{c}_i,\hat{k}_i;s) = \hat{c}_iX_i(\hat{c}_i,\hat{k}_i;s) + \int_{\hat{c}_i}^{\overline{c}_i}
	X_i(z,\hat{k}_i;s)dz\; .
\end{align}
The challenge here is to compute the integral as the allocation is not known for bid profiles other then $\hat{c}$. The allocation therein depends on how the qualities are learnt. In order to compute this integral, we adopt a sampling procedure and transformation that uses~\Cref{thm:babaioff10} similar to~\cite{BABAIOFF10}.

\begin{lemma}
\label{thm:babaioff10}
Let $\mathcal{F}:I \rightarrow [0,1]$ be any strictly increasing function that is differentiable and satisfies $inf_{z \in I}\mathcal{F}(z) = 0$ and $sup_{z \in I}\mathcal{F}(z)=1$. If $Y$ is a random variable with cumulative distribution function $\mathcal{F}$, then
\begin{align}
\label{eqn:est_integral}
\int_{I}g(z)dz = \mathbb{E}\bigg[\frac{g(Y)}{\mathcal{F}'(Y)}\bigg]\; .
\end{align}
\end{lemma}

Our self-resampling procedure is given in Algorithm \ref{alg:resampling} that returns vectors $\alpha, \beta$ based on input bids. These vectors are then used to compute the allocation and payment.
\begin{algorithm}[h!]{}
\small
\DontPrintSemicolon
\caption{Self-resampling Procedure}\label{alg:resampling}
\KwIn {bid $\hat{c}_i \in [\underline{c}_i,\overline{c}_i]$, parameter $\mu \in (0,1)$}
\KwOut {$(\alpha_i, \beta_i)$ such that $\overline{c}_i \ge \alpha_i \ge \beta_i \ge \hat{c}_i$}
\textbf{with probability} $(1-\mu)$\;
\hspace{0.2in}$\alpha_i \leftarrow \hat{c}_i$, $\beta_i \leftarrow \hat{c}_i$\;
\textbf{with probability} $\mu$\;
\hspace{0.2in} Pick $\hat{c}_i' \in [\hat{c}_i,\overline{c}_i]$ uniformly at random.\ \label{alg-step}\; 
\hspace{0.2in}$\alpha_i \leftarrow recursive(\hat{c}_i')$, $\beta_i \leftarrow \hat{c}_i'$\;
\vspace{0.1in}
\textbf{function} Recursive($\hat{c}_i$)\;
\hspace{0.2in} \textbf{with probability} $(1-\mu)$\;
\hspace{0.4in} return $\hat{c}_i$\;
\hspace{0.2in} \textbf{with probability} $\mu$\;
\hspace{0.4in} Pick $\hat{c}_i' \in [\hat{c}_i,\overline{c}_i]$ uniformly at random.\; 
\hspace{0.4in} return Recursive($\hat{c}_i'$)\;
\end{algorithm}

In order to compute the integral, we need certain properties to be satisfied that are described in Lemma \ref{lemma:prop_resampling}.
\begin{lemma}
\label{lemma:prop_resampling}
The procedure in~\Cref{alg:resampling} satisfies the following properties $\forall i \in N$:
\begin{enumerate}[leftmargin=0.4cm]
\itemsep0em 
\item{$\alpha_i(\hat{c}_i)$ and $\beta_i(\hat{c}_i)$ are non-decreasing functions of $\hat{c}_i$}\label{resampling:property1}
\item {\emph{(A)} With probability $(1-\mu)$, $\alpha_i(\hat{c}_i) = \beta_i(\hat{c}_i) = \hat{c}_i$. \\
\indent\emph{(B)} With probability $\mu$, $\overline{c}_i \ge \alpha_i(\hat{c}_i) \ge \beta_i(\hat{c}_i) > \hat{c}_i$}\label{resampling:property2}
\item {$\Prob[\alpha_i(\hat{c}_i) > a_i|\beta_i(\hat{c}_i) = \hat{c}_i'] = \Prob[\alpha_i(\hat{c}_i') > a_i]\;\ \forall a_i \ge \hat{c}_i'>\hat{c}_i$.}\label{resampling:property3}
\item {Function $\mathcal{F}(a_i,\hat{c}_i) = \Prob[\beta_i(\hat{c}_i) < a_i|\beta_i(\hat{c}_i) > \hat{c}_i] = \frac{a_i-\hat{c}_i}{\overline{c}_i-\hat{c}_i}$.}\label{resampling:property4}
\end{enumerate}
\end{lemma}
\begin{pfof}
Properties \ref{resampling:property1}, \ref{resampling:property2} are immediate from the algorithm. If $\beta_i(\hat{c}_i) = \hat{c}_i' > \hat{c}_i$, it means the algorithm has followed~\cref{alg-step} of~\cref{alg:resampling} and thus property \ref{resampling:property3} follows. Property \ref{resampling:property4} follows from the fact that distribution of $\beta_i(\hat{c}_i)$ is uniform in the interval $[\hat{c}_i,\overline{c}_i]$ conditional on the event $\beta_i(\hat{c}_i) > \hat{c}_i$
\end{pfof}
The algorithm that outputs the transformed allocation and the payment is described in Algorithm \ref{alg:transformation}.

\begin{algorithm}[h!]{}
\small
\DontPrintSemicolon
\caption{Mechanism Transformation}\label{alg:transformation}
\KwIn {$\forall i$, bids $\hat{c}_i \in [\underline{c}_i,\overline{c}_i]$, $\hat{k}_i \in [\underline{k}_i,k_i]$, parameter $\mu \in (0,1)$, allocation rule $x$}
\KwOut {Allocation rule $\tilde{x}$ and the payment rule $\tilde{t}$}
Obtain modified bids as $(\alpha,\beta) = ((\alpha_1(\hat{c}_1),\beta_1(\hat{c}_1), (\alpha_2(\hat{c}_2),\beta_2(\hat{c}_2)),\ldots,(\alpha_n(\hat{c}_n),\beta_n(\hat{c}_n))$\;
Allocate according to $\tilde{x}(\hat{c},\hat{k}) = x(\alpha(\hat{c}),\hat{k})$\;
Make payment to each agent $i$, $\tilde{t}_i(\hat{c},\hat{k}) = \hat{c}_i\tilde{x}_i(\hat{c},\hat{k}) + P_i$, where, \[
P_i =
\begin{cases}
\frac{1}{\mu}\frac{x_i(\alpha(\hat{c}),\hat{k})}{\mathcal{F}_i'(\beta_i(\hat{c}_i),\hat{c}_i)},\ \text{if}  \displaystyle \beta_i(\hat{c}_i) > \hat{c}_i\\
0,\ \text{otherwise.}
\end{cases}
\]
\end{algorithm}

\begin{pf}{Theorem}{\ref{lemma:online_payment}}
We will prove that the transformed mechanism in Algorithm \ref{alg:transformation} satisfies all the properties in Theorem \ref{thm:online_bic} when the input allocation rule is \MakeLowercase{\PRIN} and thus is stochastic BIC and IR. Transformed allocation and payment rule are denoted by $\tilde{x}$ and $\tilde{t}$ respectively. We denote $\tilde{X}_i(\hat{c}_i,\hat{k}_i;s)$ $= \mathbb{E}_{b_{-i},\alpha}[x_i(\alpha(\hat{c}),\hat{k};s)]$ as the expected allocation with the expectation taken over randomization of the algorithm and bid profile of other agents. Similarly, we denote $\tilde{T}_i(\hat{c}_i,\hat{k}_i;s)$ $ = \mathbb{E}_{b_{-i},\alpha,\beta}[t_i(\alpha(\hat{c}),\beta,\hat{k};s)]$. For all reward realizations $s$, we will prove two properties: (1) Allocation rule $\tilde{X}$ is monotone in terms of costs, and (2) the expected payment rule $\tilde{T}$ satisfies~\cref{eq:truthfulness}.

 The monotonicity of allocation rule $\tilde{X}$ follows from the monotonicity of $x$ (Lemma \ref{lemma:online_monotone}) and the monotonicity property \ref{resampling:property1} of Algorithm \ref{alg:resampling} (Property 1, Lemma \ref{lemma:prop_resampling}). 

We now prove that $\mathbb{E}_{b_{-i},\alpha,\beta}[P_i] = \int_{\hat{c}_i}^{\overline{c}_i}\tilde{X}_i(\hat{k}_i,z;s)dz$, where the expectation is taken over bids of other players as well as over the randomization of the \Cref{alg:transformation}. 
{\allowdisplaybreaks
\begin{align*}
&\mathbb{E}_{b_{-i},\alpha,\beta}[P_i] \\
&= \mathbb{E}_{\beta_i}\mathbb{E}_{b_{-i},\alpha|\beta_i}[P_i] \tag*{($P_i$ does not depend on $\beta_{-i}$)}\\
&= \mathbb{P}(\beta_i > \hat{c}_i)\mathbb{E}_{\beta_i|\beta_i > \hat{c}_i}\mathbb{E}_{b_{-i},\alpha|\beta_i}[P_i] \tag*{($P_i=0$ if $\beta_i = \hat{c}_i$)}\\
&= \mu\mathbb{E}_{\beta_i|\beta_i > \hat{c}_i}\mathbb{E}_{b_{-i},\alpha|\beta_i}\bigg[\frac{x_i(\alpha(\hat{c}),\hat{k};s)}{\mu \mathcal{F}_i'(\beta_i(\hat{c}_i),\hat{c}_i)}\bigg] \tag*{(Property 2 of Lemma \ref{lemma:prop_resampling})}\\
&= \mathbb{E}_{\beta_i|\beta_i > \hat{c}_i}\frac{1}{\mathcal{F}_i'(\beta_i,\hat{c}_i)}\mathbb{E}_{b_{-i},\alpha}[x_i(\alpha_i(\beta_i),\alpha_{-i}(\hat{c}_{-i}),\hat{k};s)]\tag*{(Property 3 of Lemma \ref{lemma:prop_resampling})}\\
&=\mathbb{E}_{\beta_i|\beta_i > \hat{c}_i}\frac{\tilde{X}_i(\beta_i,\hat{k}_i;s)}{\mathcal{F}_i'(\beta_i,\hat{c}_i)}\\
&= \int_{\hat{c}_i}^{\overline{c}_i}\tilde{X}_i(z,\hat{k}_i;s)dz \tag*{(Lemma \ref{thm:babaioff10})}
\end{align*}
}
 We also have,
\begin{align*}
\rho_i(\overline{c}_i,\hat{k}_i;s) &= \tilde{T}_i(\overline{c}_i,\hat{k}_i;s) - \overline{c}_i\tilde{X}_i(\overline{c}_i,\hat{k}_i;s) \tag*{(\cref{eq:rho_utility})}\\ 
&=\overline{c}_i\tilde{X}_i(\overline{c}_i,\hat{c}_{-i},\hat{k};s) -\int_{\overline{c}_i}^{\overline{c}_i}\tilde{X}_i(z,\hat{k}_i;s)dz- \overline{c}_i\tilde{X}_i(\overline{c}_i,\hat{k}_i;s)\\
&=0
\end{align*}
\noindent Thus, $\rho_{i}(\hat{c}_i,\hat{k}_i;s) = \rho_{i}(\bar{c_i,}\hat{k}_i;s) + \int_{\hat{c}_i}^{\bar{c}_i}X_i(z,\hat{k}_i;s)dz$. Since the allocation rule is monotone in capacity, $\rho_{i}(b_i;s)$ non-negative, and non-decreasing in $\hat{k}_i$, $\forall s$ and  $\forall\hat{c}_i \in [\underline{c}_i,\bar{c}_i]$.
\end{pf}

% \begin{theorem}
% \label{thm:online_main}
% Any \MakeLowercase{\PRIN} allocation rule can be implemented as a stochastic BIC and IR mechanism.
% \end{theorem}
% \begin{pfof}
% From \Cref{lemma:online_payment}, all the conditions in \Cref{thm:online_bic} are satisfied for each reward realization, thus, resulting in a stochastic BIC and IR mechanism.
% \end{pfof}

\subsection{\CUCB: A Learning Mechanism}
\noindent With the necessary machinery established, we now present the learning mechanism  given in \Cref{alg:online_mechanism}. Mechanism \CUCB procures one unit at a time, learns the quality and makes the allocation similar to \CHOPT on the basis of learnt qualities so far. The payment is computed with the help of transformed mechanism given in \Cref{alg:transformation}. 
\begin{algorithm}[h!]{}
\small
\DontPrintSemicolon
\caption{\CUCB Mechanism}\label{alg:online_mechanism}
\KwIn {$\forall i \in N$, bids $\hat{c}_i \in [\underline{c}_i,\overline{c}_i]$, $\hat{k}_i \in [\underline{k}_i,k_i]$, parameter $\mu \in (0,1)$, Reward parameter $R$}
\KwOut {A mechanism $\mathcal{M} = (x,t)$ }
$\forall i \in N$, $\hat{q}_i^+ = 1$, $\hat{q}_i^- = 0$, $n_i = 1$\;
Obtain modified bids as $(\alpha,\beta)$\\ $= ((\alpha_1(\hat{c}_1),\beta_1(\hat{c}_1), \ldots,(\alpha_n(\hat{c}_n),\beta_n(\hat{c}_n))$ using \cref{alg:resampling}\;
Allocate one unit to all agents and estimate empirical quality $\hat{q}$\;
$\hat{q}_i = \tilde{q}_i(i)/n_{i}$, $\hat{q}_i^+ = \hat{q}_i + \sqrt{\frac{1}{2n_{i}} ln(t)}$\;
\For{$t=n$ to $L$}{
Compute $H_i = \alpha_i + \frac{F_i(\alpha_i|\hat{k}_i)}{f_i(\alpha_i|\hat{k}_i)}$\;
Let $i = \argmax_{\{j s.t. k_j > n_j\}} R\hat{q}_j^+ - H_j$ and $\hat{G}_i = R\hat{q}_i^+ - H_i$\;
\If {$\hat{G}_j > 0$}
{Procure the unit from agent $i$ and update $\hat{q}_i$\;
 $\hat{q}_i^+ = \hat{q}_i + \sqrt{\frac{2}{n_{i}} ln(t)}$
 }
\Else{break $\backslash\backslash$ Don't allocate future units to anyone}}
Make payment to each agent $i$, $\tilde{T}_i = \hat{c}_in_i + P_i$, where, 
\begin{align*}
P_i =
\begin{cases}
\frac{1}{\mu}n_i(\overline{c}_i-\hat{c}_i),\ \text{if}  \displaystyle \beta_i > \hat{c}_i\\
0,\ \text{otherwise.}
\end{cases}
\end{align*}
\end{algorithm}

\begin{theorem}
\label{thm:chucb_bic}
\CUCB is stochastic BIC and IR.

\end{theorem}
\begin{pfof}
We first prove that the allocation rule produced by \CUCB mechanism is well-behaved. At every time, the mechanism allocates the unit to an agent with highest value of $\hat{G}_i$. The value of $\hat{G}_i$ only depends on learnt quality so far. It is monotone in terms of cost due to regularity assumption and monotonicity property of \Cref{alg:resampling}. Thus Property $1$ of \MakeLowercase{\PRIN} is satisfied. If an agent reduces his capacity then he might lose an allocation since no agent is allocated more then his bid capacity thus satisfying property $3$. The allocation rule also satisfy property $2$ (IIA) since the allocation is made to the agent with highest $\hat{G}_i$ and if agent $i$ changes his bid then it will not affect the $\hat{G}_i's$ of other agents. Since the payment structure follows from \cref{alg:transformation}, and conditions of \Cref{thm:online_bic} are also satisfied and thus the resulting mechanism is stochastic BIC and IR.
\end{pfof}

\section{Simulations}
\label{sec:simulations}
\noindent In \Cref{sec:online}, we have presented a learning mechanism \CUCB, which embeds \CHOPT. We have theoretically established the optimality of \CHOPT when the qualities of the agents are known. A detailed regret analysis of our learning mechanism \CUCB  will be quite involved and forms an interesting future direction.  We instead evaluate the performance of our learning mechanism via simulations.

In the simulations, we compare the expected utility per unit given by \CUCB against the optimal benchmark \CHOPT which is fully aware of underlying quality. Another good benchmark to compare against is an $\varepsilon-$separated mechanism. An $\varepsilon-$separated mechanism allocates $\varepsilon L$ units to all the agents irrespective of their bids. Based on the observed realization, the learned qualities in these rounds are used to find the allocation and payments in $(1-\varepsilon)L$ future rounds using \CHOPT and also qualities are not updated further. It is easy to verify that an $\varepsilon-$ separated mechanism is BIC and IR.
% \end{pfof}

\begin{figure}[h]
\centering
 \includegraphics[width=2.8in]{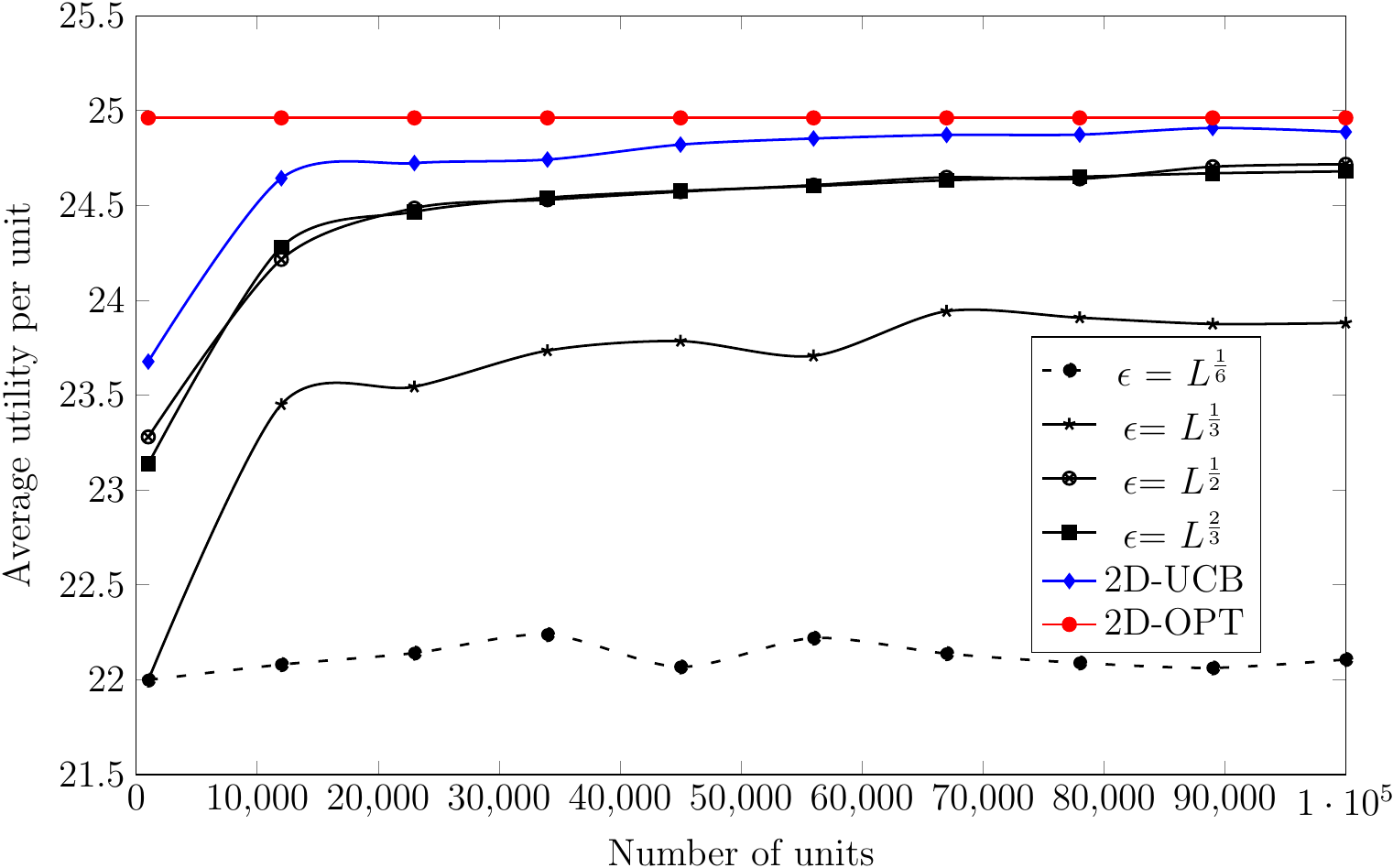}
\caption{Comparative study of average utility per unit}
 \label{fig:avgutility}
\end{figure}

For the simulations, the number of units of the item ($L$), which the auctioneer wishes to procure, is chosen at first as $10^3$ and subsequently at nine other linearly spaced steps from $10^3$ to $10^5$. We choose a pool of five agents($N$). A unit procured from an agent $i$ yields a Bernoulli reward with mean $q_i$ drawn uniformly from the interval $[0.5,1]$. The private types of the agents are independently distributed and the costs are drawn uniformly from $[0,1]$. The cost and capacity are chosen to be independently distributed and therefore the setup meets regularity.  The capacity is a positive integer drawn with equal probability in the range with upper limit as $L$ and lower limit large enough to meet the uniform exploration. For this type distribution, it can be shown that the virtual cost function for an agent $i$ is $H_i = 2c_i$ by simple computation. For the $\varepsilon$-separated mechanisms, we choose the number of exploration rounds as $\{L^{1/6},L^{1/3},L^{1/2},L^{2/3}\}$. A Bernoulli reward 1 of a procured instance yields a reward of $R=30$ to auctioneer. The performance measure used is the expected average utility per unit obtained by the auctioneer plotted as a function of the number of units. To estimate the expected average utility, 200 independent samples are drawn from the type distribution; for each such sample the number of units required to be procured is varied; at each value of $L$ multiple instances($100$) of reward realization is drawn from the true underlying quality. As $L$ is varied, the capacity is suitably scaled yielding a constant average utility for the benchmark as shown in ~\cref{fig:avgutility}. We choose $\mu=0.1$ for \CUCB.

The simulations indicate that all the mechanisms yield average utilities per unit which asymptotically converge to \CHOPT. The performance of \CUCB however is superior in the sense that it approaches \CHOPT faster.

\section{Conclusion}
\label{sec:conclusion}
\noindent We have studied a class of mechanisms which yield a stochastic reward to the  auctioneer following an allocation to an agent.  We have presented optimal learning mechanisms which truthfully elicit multiple private types. A corresponding welfare maximizing version follows directly from the ideas presented in this paper. It would be interesting to study a setting where the allocation is over a subset of agents rather than a single agent. A complete characterization of a learning algorithm in this space is still open as we have provided only sufficient conditions.  Also, a theoretic lower bound on regret would be interesting.
\newpage
\printbibliography
\end{document}